\documentclass[a4paper,11pt]{article}
\usepackage{pos}
\usepackage{gensymb}
\usepackage{physics}

\title{A reconstruction procedure for near horizon extensive air showers based on radio signals}
 \ShortTitle{EAS Radio Reconstruction}

\author*[a]{Valentin Decoene}
\author[b, d]{Olivier Martineau-Huynh}
\author[c]{Mat\`ias Tueros}
\author[d]{Simon Chiche}

\affiliation[a]{Department of Physics, The Pennsylvania State University, University Park, Pennsylvania 16802, USA}

\affiliation[b]{Sorbonne Universit\'e, Universit\'e Paris Diderot, Sorbonne Paris Cit\'e, CNRS/IN2P3, LPNHE, Paris, France}

\affiliation[c]{Instituto de F\'isica La Plata - CONICET, Argentina}

\affiliation[d]{Sorbonne Universit\'e, CNRS, UMR 7095, Institut d'Astrophysique de Paris, 98 bis bd Arago, 75014 Paris, France}

\emailAdd{decoene@psu.edu}

\abstract{Very inclined extensive air showers (EAS), with both down-going and up-going trajectories, are particularly targeted by the next generation of extended radio arrays, such as GRAND. Methods to reconstruct the incoming direction, core position, primary energy and composition of showers with these specific geometries, remain to be developed.
\\Towards that goal, we present a new reconstruction procedure based on the arrival times and the amplitudes of the radio signal, measured at each antenna station. This hybrid reconstruction method, harnesses the fact that the emission is observed, at the antenna level, far away from the emission region, thus allowing for a point-like emission description. Thanks to this assumption, the arrival times are modelled following a spherical wavefront emission, which offers the possibility to reconstruct the radio emission zone as a fixed point along the shower axis. From that point the amplitude distribution at the antenna level is described through an Angular Distribution Function (ADF) taking into account at once all geo-magnetic asymmetries and early late effects as well as additional signal asymmetries featured by very inclined EAS. This method shows promising results in terms of arrival direction reconstruction, within the $0.1\degree$ range, even when taking into account experimental uncertainties, and interesting potential for the energy reconstruction and primary composition identification.}

\FullConference{37$^{\rm{th}}$ International Cosmic Ray Conference (ICRC 2021)\\
		July 12th -- 23rd, 2021\\
		Online -- Berlin, Germany}

\begin{document}
\maketitle

%----------------------------------------------------------------------
\section{Motivations}
Near horizon extensive air showers (EAS) are particularly targeted by the next generation of extended radio arrays~\cite{GRAND}. The main motivation comes from the large radio footprint resulting from the intersection between the radio beam and the ground. This feature enables the possibility to deploy sparse arrays over very large areas, hence meeting two interesting criteria: a large detector effective area and a substantial reduction of costs.
Such an approach enhances the detection of both up-going and down-going extensive air showers, hence the quest of ultra high energy neutrinos and ultra high energy cosmic rays.

For technical and technological reasons, the study of EAS has been primarily focused on vertical events. Therefore the methods and tools for the study of very inclined extensive air showers remains nowadays largely underdeveloped.

In the perspective of filling this gap, we developed a new reconstruction method. The procedure relies on a hybrid approach where both the information on the arrival time and the amplitude of the signal is used to reconstruct step by step: emission point, arrival direction and amplitude. The location of the emission point shows potential for its  later use as a primary discriminator, and the amplitude as a proxy for the determination of the primary energy.

In the next sections we details how exactly each information is exploited.

%----------------------------------------------------------------------
\section{Method}
\label{sec:method}
The reconstruction procedure relies on the combination of information from both the arrival time and the amplitude of the radio signal at each antenna station. Specifically, the detected arrival times allow for a reconstruction of the emission point under the assumption that the wavefront emission is spherical, or more precisely seen as spherical by the detector, as detailed in subsection~\ref{subsec:wavefront}. 
From that reconstructed emission point and with the description of the amplitude pattern measured on ground, we can reconstruct the position of the shower axis, hence solving the geometry of the EAS. This description takes into account all asymmetry features and describes in a phenomenological approach how the signal is diluted through its propagation and direction of observation, as shown in subsection~\ref{subsec:amplitude}. Finally by combining these two steps, the reconstruction procedure allows to extract directly the emission point and direction, and indirectly the energy and composition of the extensive air shower.

%-------------------------------------------------
\subsection{Arrival times and emission point}
\label{subsec:wavefront}
%From classical electrodynamics considerations, it is easy to understand that key information about the EAS such as the geometry (direction and location) or the content (energy and composition) can be extracted from the emitted radio signal.
%Following these considerations, the arrival times measured at the antenna level imprints the wavefront of the induced radio emission by the EAS, and is expected in the general case to be strongly dependant on the EAS direction. 
%Therefore an accurate modelling of the wavefront is required to achieve the reconstruction of the EAS direction from the radio signals detected at the antennas level.

To reconstruct the EAS direction from the detected radio signals, accurate modelling of the wavefront is required.
Phenomenological models have been envisioned to that end and validated through empirical studies on simulated and real data~\cite{corstanje}.
\begin{figure}[htbp]
\centering
\includegraphics[width=0.99\linewidth]{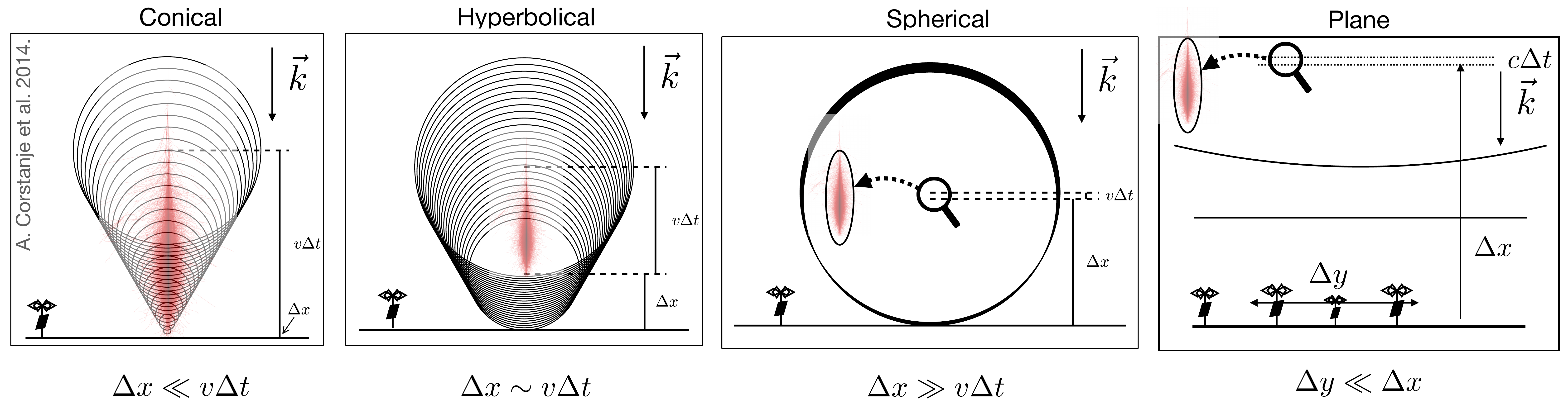}
\caption{Sketch of the the four typical wavefront shapes expected from the radio emission of EAS, depending on the configuration between emission area $\Delta x$, propagation distance $v\Delta t$ and detector extension $\Delta y$. From left to right we expect: a conical shape when the emission zone is very near-by and extended; in the intermediate case, a hyperbolic shape, when both propagation distance and emission zone are comparable; inversely, a spherical shape when the emission zone is very small compared to propagation distance. And finally, a flat shape similar to a plane wave emission, when the detector extension is too small compared to the curvature of the emission.}
\label{fig:wavefront_models}
\end{figure}
Fig.~\ref{fig:wavefront_models} depicts these models for four typical configurations of EAS geometry and detector. Even if in general  the wavefront from the shower radio emission is close to a hyperbola~\cite{Apel_2014,corstanje}, the location of the emission zone  relative to the detector array and its capacity to sample precisely the emission wavefront are key parameters in the interpretation of the observed radio wavefront.

In this study, this interpretation has been tested in the regime of very inclined EAS and a sparse array. Several inclined EAS were simulated using the ZHAireS software~\cite{Alvarez_Mu_iz_2012}, covering zenith angles ranging from $\theta=63\degree-87.1\degree$, energies covering the range of $E=0.02-4$\,EeV and azimuth towards North, East and South. The simulated antennas were positioned into $7$ planes with "star" shape patterns and located perpendicular to the shower axis along the propagation direction at specific distances from $X_{\rm max}$ in the range $17-200$\,km.
These configurations allow for an efficient sampling of the wavefront curvature and its evolution with propagation.
The wavefront arrival times were defined as the maximum of the Hilbert envelope of the electric field simulated at each antenna. Arrival times turned out to be independent of the filtering, at least in the range $50-200$\,MHz.
Analysis of these simulations showed that the wavefront shape is independent not only of the primary nature and its energy, but also of the shower zenith angle, and that it only depends on the propagation distance. The wavefront shape was computed within each antenna plane by subtracting from the arrival time of each antenna $t_i$, the pure propagation delay $l_{\rm emission} \frac{n}{c}$, which depends on the refractive index $n$, speed of light $c$ and distance from the $X_{\rm max}$ neighbourhood $l_{\rm emission}$. Hence giving a measure of the curvature of the wavefront shape at a given propagation distance, for different lateral antenna positions, in terms of residual time delays $t_{\rm delay, i} = t_i - l_{\rm emission} \frac{n}{c}$.
\begin{figure}[htbp]
\centering
\includegraphics[width=0.49\linewidth]{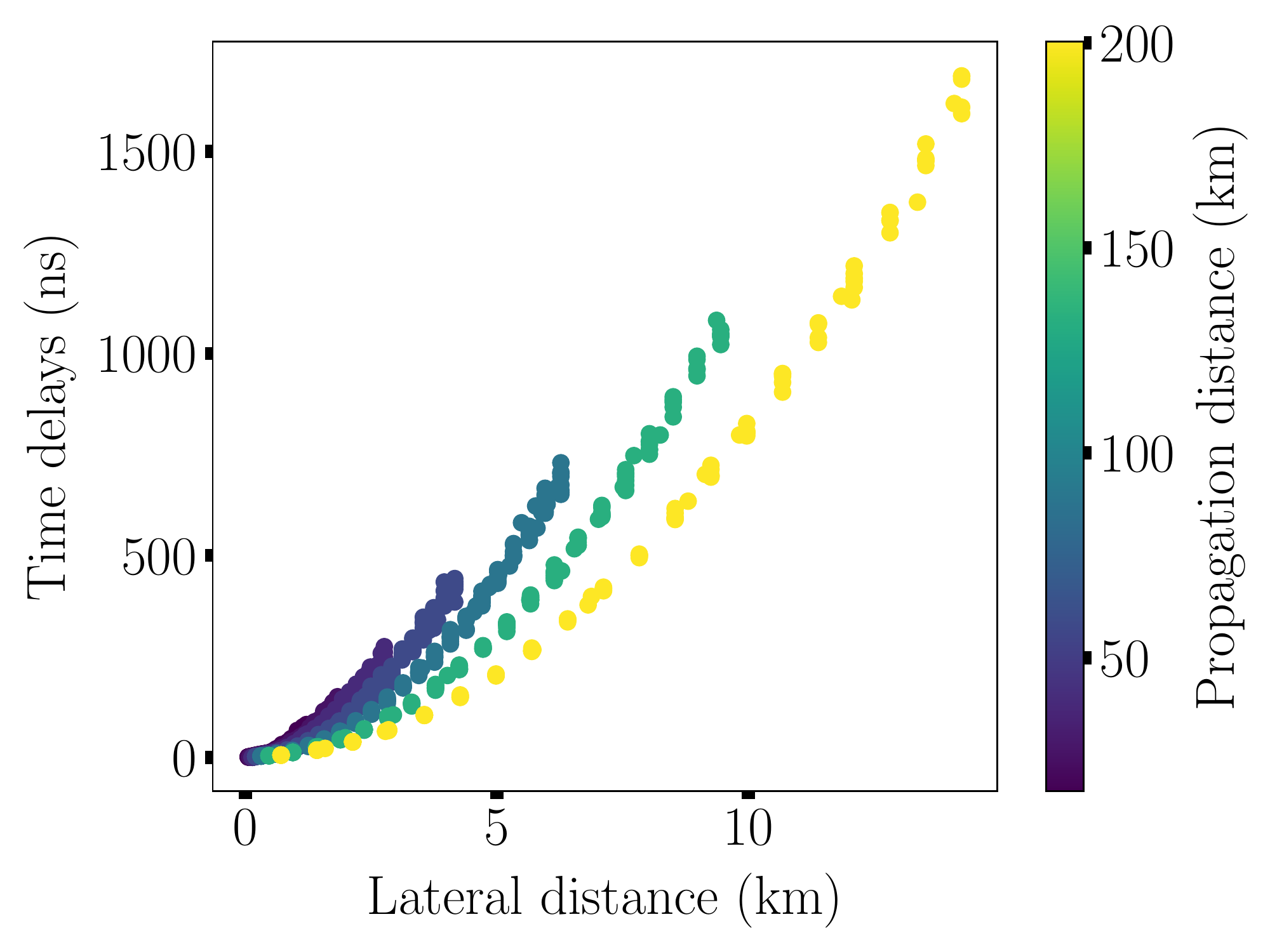}
\includegraphics[width=0.49\linewidth]{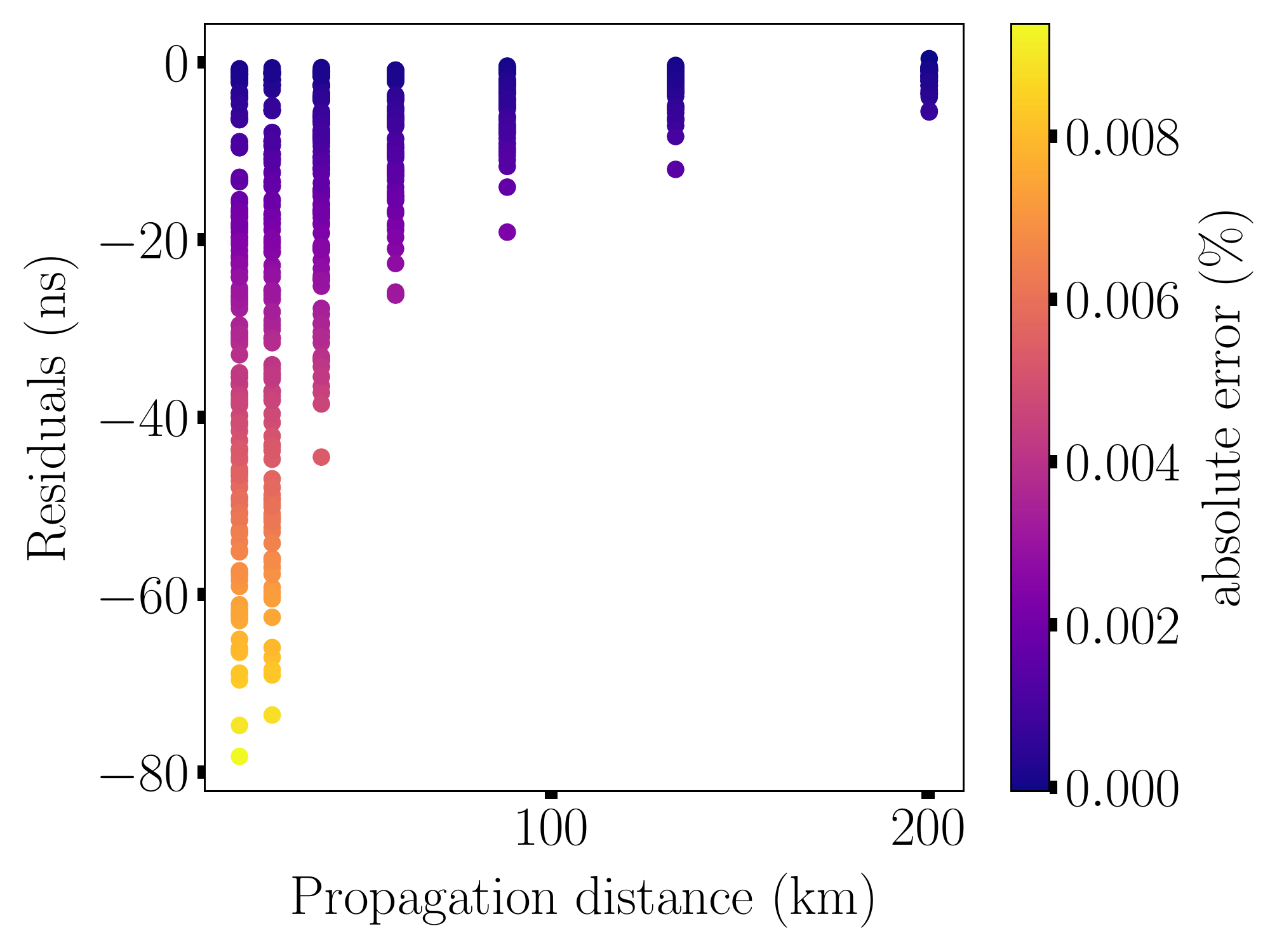}
\caption{{\it Left:} time delays of the arrival time with respect to the propagation plane, as a function of lateral distance and propagation distance in colour-code. {\it Right:} residual of the spherical model against the arrival times, as a function of propagation distance and absolute error in colour-code. For a shower with direction $\theta=87.2\degree$, $\phi=0\degree$ and energy $E=4$\,EeV. Antennas are placed within $7$ planes located from $17$\,km to $200$\,km from $X_{\rm max}$.}
\label{fig:sphere_wavefront}
\end{figure}
An example of this curvature is displayed in Fig.\ref{fig:sphere_wavefront}, left. As expected the time delays increase with lateral distance due to the curvature of the wavefront but the overall curvature effect tends to flatten with propagation distance. This flattening of the arrival times is also expected as a geometrical expansion of the central part of wavefront shape with respect to the shower axis. Therefore at large distances from the emission region, which is expected to be located close to $X_{\rm max}$, a spherical curvature for the wavefront shape is expected.

The comparison of the arrival time at each antenna with the expected arrival time for a "spherical" curvature model is illustrated in Fig.\ref{fig:sphere_wavefront} right. The curvature model is strictly speaking a spheroid since it takes into account the effective refractive index along the line of sight of each antenna, as shown in the following analytical expression 
$ t_{\rm sph, i} = \frac{n_i}{c} \sqrt{l_i + r_i}$,
where $l_i$ and $r_i$ corresponds respectively to the longitudinal and lateral positions of the antenna, and $n_i$ is the effective refractive index along its line of sigh to the emission point of the EAS.

From Fig.~\ref{fig:sphere_wavefront} two main conclusions can be drawn: (i) the wavefront shape tends to a spherical shape with increasing propagation distance, (ii) the residuals are on average below any current experimental resolution for propagation distances above $\sim50$\,km. Therefore the wavefront is totally undistinguishable between a sphere or a more complex shape. Furthermore one can notice the good agreement between the model and the simulations with an absolute error below $0.002\%$.

An important implication of this result is that it is possible to identify an emission point associated to the geometrical shape of the spherical wavefront. This emission point is by construction located along the shower axis and can be used in a reconstruction of the shower axis.

The determination of the radio emission point-source presented above can be associated to the reconstruction of the shower axis presented in the next section. From the combination of both, an estimation of the direction of origin of the primary particle, together with its nature and shower energy can be inferred. This is detailed in section~\ref{subsec:energy_composition}. 

%-------------------------------------------------
\subsection{Angular Distribution Function (ADF)}
\label{subsec:amplitude}
Unlike for the study of the arrival time wavefront, here the description of the amplitude pattern within the radio footprint is purely empirical (and is only motivated by reconstruction goals).
As already mentioned, the ADF model describes the amplitude distribution inside the footprint of the radio emission thanks to a frame located at the emission point and angular coordinates. Concretely, any point is described as a function of an angle $\omega$ measured from the emission point with respect to the shower axis, a longitudinal propagation distance $l$ along the shower axis from the emission point, and few additional parameters described below.
The model can be written as
\begin{align} \label{eq:adf_model}
f^{\rm ADF}\qty(\omega, \eta, \alpha, l ; \delta \omega, \mathcal{A}) = \frac{\mathcal{A}}{l} f^{\rm GeoM}\qty(\alpha, \eta, \mathcal{B})\ f^{\rm Cerenkov}\qty(\omega,  \delta \omega) \ ,
\end{align}
where $\mathcal{A}$ is a free parameter adjusting the amplitude of the signal, and $f^{\rm GeoM}\qty(\alpha, \eta, \mathcal{B})$ describes the well-known geomagnetic asymmetry and is given by
\begin{align} \label{eq:geom_model}
f^{\rm GeoM}\qty(\alpha, \eta, \mathcal{B}) = 1 + \mathcal{B}\sin{\qty(\alpha)}^2 \cos{\qty(\eta)} \ ,
\end{align}
where $\mathcal{B} \sim 0.005$ is the geomagnetic asymmetry strength (adjusted from simulations and in the range of expected value), $\alpha$ is the angle between the shower direction and the magnetic field, $\eta$ is the antenna angle with respect to the $\vec{k} \times \vec{B}$ axis, in the well-known shower-plane~\cite{Apel_2014}.

The Cerenkov pattern is modelled thanks to a Lorentzian distribution, since for inclined showers it has been found to describe best the sharp rise observed close to the Cerenkov angle. The expression of $ f^{\rm Cerenkov}\qty(\omega, \delta \omega)$ is given by
\begin{align} \label{eq:cerenkov_model_adf}
 f^{\rm Cerenkov}\qty(\omega, \delta \omega) = \frac{1 }{1 + 4 \qty[\frac{\qty(\tan{\qty(\omega)}/\tan{\qty(\omega_{\rm C})})^2 - 1}{\delta \omega}]^2} \ ,
\end{align}
where $\omega_{\rm C}$ is the Cerenkov angle computed from a model presented below and $\delta \omega$ is a free parameter describing the width of the Cerenkov cone.

All the variables used in this model can be written explicitly as a function of the shower direction $\vec{k}$
\begin{align}
\omega_{i} &= \acos{\qty(\vec{k} \cdot \vec{x}_i)} \ ,\  l_{i} = \vec{k} \cdot \vec{x}_i \ , \ \eta_{i} = \atan{\qty(y_i^{\rm sp} / x_i^{\rm sp})} \ , \ \alpha = \acos{\qty(\vec{k} \cdot \vec{B})} \ ,
\end{align}
where $\vec{x}_i$ is the antenna position with respect to the shower source and $x_i^{\rm sp}$ , $y_i^{\rm sp}$ are the antenna coordinates in the shower plane defined as the projections on $\vec{k} \times \qty(\vec{k} \times \vec{B})$ and $\vec{k} \times \vec{B}$. Note however, that the ADF model does not depend on the shower core position, which makes it usable as well for up-going EAS with no core on the ground. Therefore, it provides a great handle for the reconstruction of the shower direction of propagation.

\paragraph{Cerenkov angle computation}
The Cerenkov angle corresponds to the angle between the shower axis and the well-known Cerenkov ring observed in the radio footprint on ground.
From a simulation study it has been found that in the case of very inclined EAS and realistic atmosphere model, the Cerenkov ring was no longer symmetrical by rotation around the shower axis~\cite{PhDThesis, schlter2020refractive}. Consequently the angle where to expect the sharp signal rise depends on the shower geometry and observer position, for a given atmosphere model. The effect is likely to originate from the combination of very far away emission paths and varying refractive index. Even thought no ray bending effect is included in the simulations, the time compression effect of the radio Cerenkov effect remains at play in the simulation, hence the shortest path for the radio signal becomes un-symmetrical for inclined showers. The effect remains too small to be noticed for standard showers but in our regime it cannot be ignored. A detailed computation of this effect is detailed in~\cite{PhDThesis}, and is taken into account in the ADF model.

For brevity reasons the validation of the ADF model is not detailed here but can be found in~\cite{PhDThesis}. The preliminary results presented below will vouch for it.

%----------------------------------------------------------------------
\section{Preliminary results}
\label{sec:results}
In practice, the reconstruction procedure follows three steps: first a plane wavefront reconstruction is used on the arrival times, in order to reduce the parameter space; then the emission point is determined thanks to the fitting of the arrival times with the spherical wavefront reconstruction; and finally, the direction is obtained through the fitting of the amplitudes with the ADF reconstruction. 
Details regarding the optimisation and the treatment of calibration errors and systematics are not considered here.

Three experimental scenarii are investigated to mimic the experimental uncertainties in~\cite{PhDThesis} but only the most conservative case is presented here. In this scenario, a random gaussian error with a standard deviation of $\sigma_{\rm t} = 5$\,ns is added to the trigger times and the signal amplitudes are randomised with a gaussian distribution of $\sigma_{\rm A} = 20\%$, corresponding to a conservative scenario for the amplitude calibration.

The accuracy of the reconstruction of the arrival direction is measured in terms of angular distance $\psi$ between the true direction ($\phi_{\rm true}$, $\theta_{\rm true}$) and the reconstructed  ($\phi_{\rm rec}$, $\theta_{\rm rec}$) through the relation
$
\cos{\qty(\psi)} = \cos{\qty(\theta_{\rm rec})} \cos{\qty(\theta_{\rm true})} + \cos{\qty(\phi_{\rm rec} - \phi_{\rm true})} \sin{\qty(\theta_{\rm true})} \sin{\qty(\theta_{\rm rec})}
$.

%-----------------------
\subparagraph{Layout and simulation set:}
Preliminary results are detailed in this section for the case of GRANDProto 300, a $\sim300$\,km$^2$ engineering array, planned in the staged approach of the GRAND experiment~\cite{GRAND}. It is designed for the detection of very inclined EAS (from $60\degree$ to $90\degree$) initiated by cosmic rays in the energy regime from $10^{16.5}-10^{18}$\,eV.

\begin{figure}[htbp]
   \centering
   \includegraphics[width=0.32\linewidth]{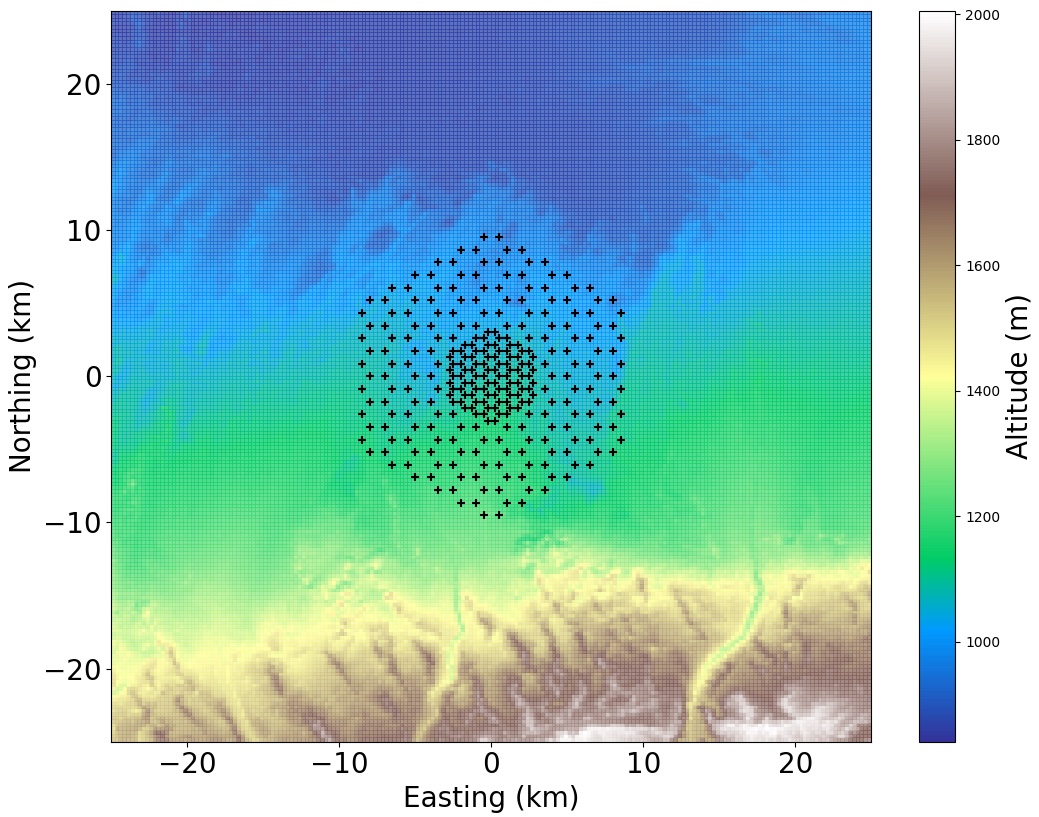}
   \includegraphics[width=0.33\linewidth]{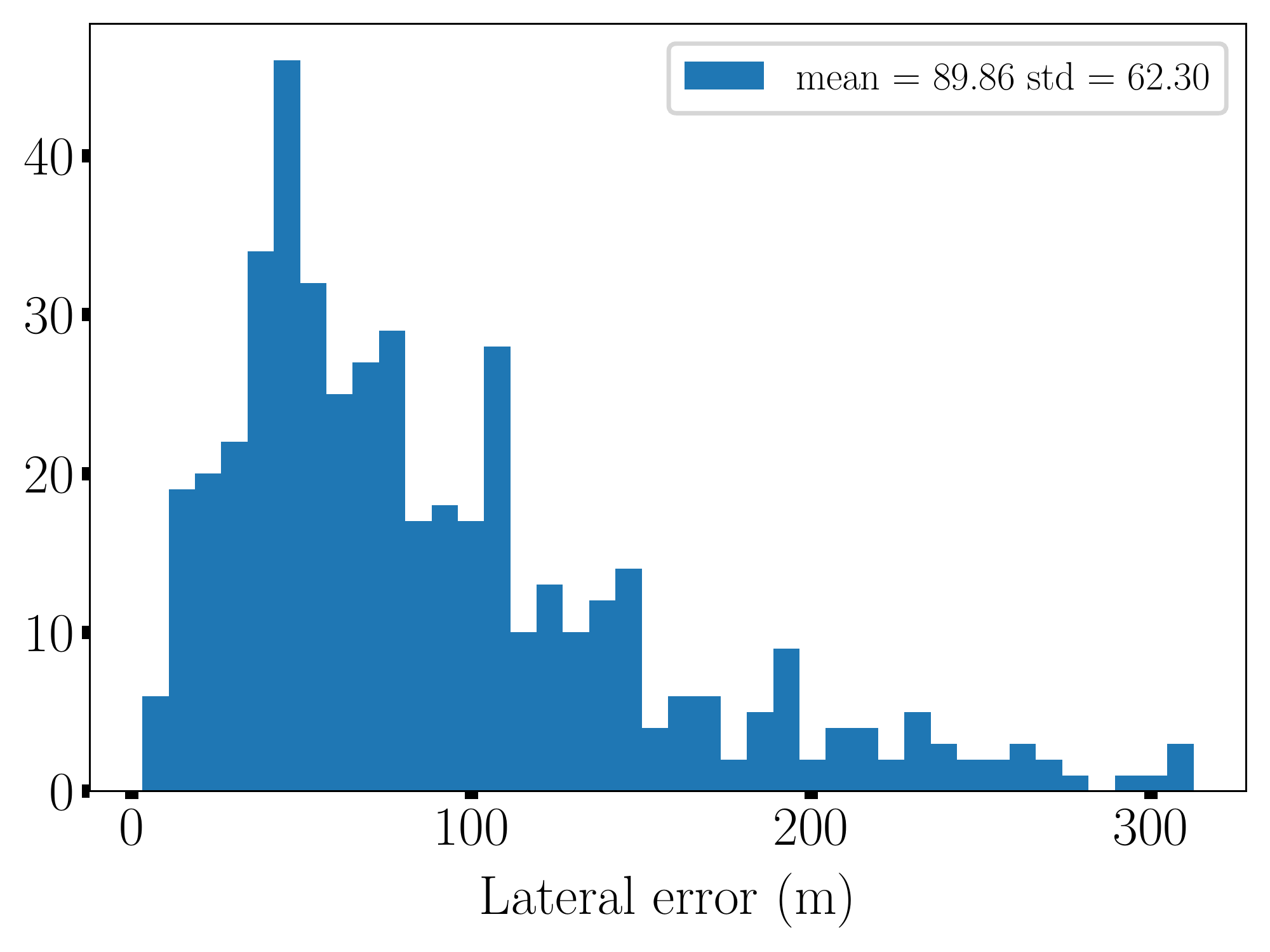} %left
      \includegraphics[width=0.33\linewidth]{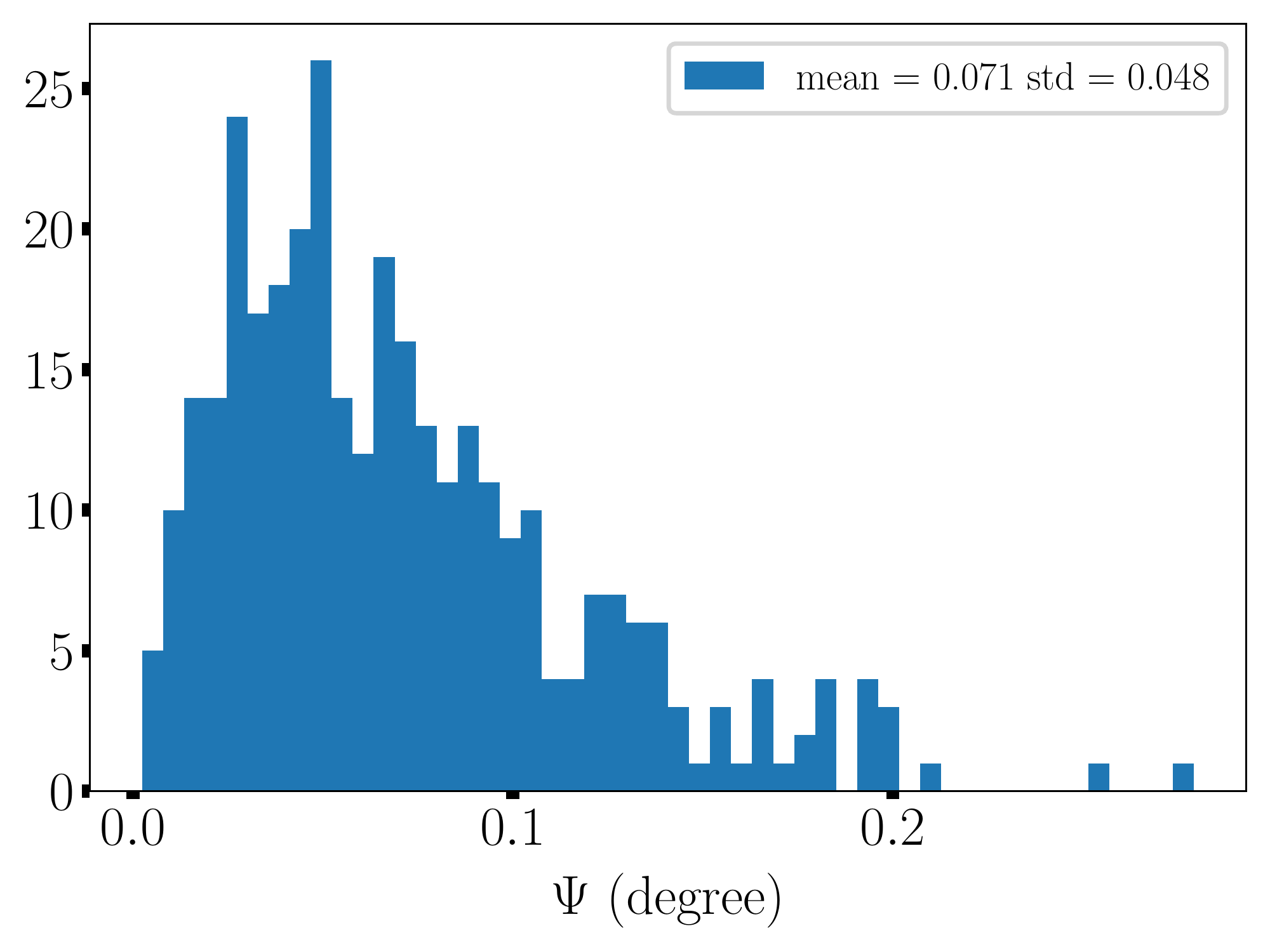}        %right
     \caption{{\it Left:} Layout of a GP300-like array taking into account the topography at one of the selected site, see text.
     {\it Middle:} Distributions of the lateral error on the reconstructed position of the shower axis. {\it Right:} Distributions of the angular distances $\psi$ between the true directions and the reconstructed ones.  For a conservative noise scenario and the GP300-like layout. The mean angular resolution reaches down to $\sim 4'$.}
   \label{fig:GP300_recons_lat_psi}
\end{figure}

Fig.~\ref{fig:GP300_recons_lat_psi} left, displays the GP300-like layout with the real topography of one of the candidate sites in Western China. The array is characterised by a hexagonal layout pattern of $215$ antennas with a $1$\,km spacing and an infill of $72$ antennas with a $500$\,m step, allowing for the detection of EAS at an energy lower than the usual values targeted by GRAND. The simulation set consists in about $1500$ simulated EAS, induced by Protons, Iron nuclei and Gamma rays with energies ranging between $0.1$\,EeV and $3.98$\,EeV with logarithmic bins, and the arrival directions are set with a random azimuth angle ($\phi$) (comprised between $0\degree$ to $360\degree$) and a zenith angle ($\theta$) between $63\degree$ and $88.2\degree$, with logarithm bins in $1/\cos{\qty(\theta)}$. The shower core positions are randomly drawn over an area larger than the array.

%%

%-------------------------------------------------
\subsection{Reconstruction of the emission-point and arrival-direction}
\label{subsec:emission_point_arrival_direction}
Fig.~\ref{fig:GP300_recons_lat_psi} middle, shows the histograms of the lateral error on the reconstruction of the emission point, with respect to the shower axis position. The reconstructed lateral position presents on average an error of about $90$\,m with a standard deviation of about $62$\,m, and does not depend on the composition of the primary particle. 
A value to be compared to the tens of kilometers up to hundreds of kilometers from the location of the emission source for very inclined EAS.
The right panel of Fig.~\ref{fig:GP300_recons_lat_psi} displays the angular resolution achieved on the arrival direction thanks to the combination of the ADF fit on top of the determination of the emission point, in particular its lateral position. As expected the achieved resolution is below the $0.1\degree$, reaching on average about $0.07\degree\sim4'$, with standard deviation around $0.05\degree\sim3'$. This accuracy is reached after applying quality cuts on convergence criteria after the minimisation and with a requirement of at least $10$\,triggered antennas within each footprint.

%-------------------------------------------------
\subsection{Energy and Composition reconstruction}
\label{subsec:energy_composition}
Thanks to the ADF model, it is possible to reconstruct the EAS electromagnetic energy from the primary, and in a second step constrain the mass composition.
To demonstrate this, a proof of principle is achieved on $\sim30,000$ simulations done over star-shaped antenna arrays deployed on a flat ground and centered around the shower core. In future developments this method will be applied to realistic arrays.

\begin{figure}[htbp]
   \centering
      \includegraphics[width=0.32\linewidth]{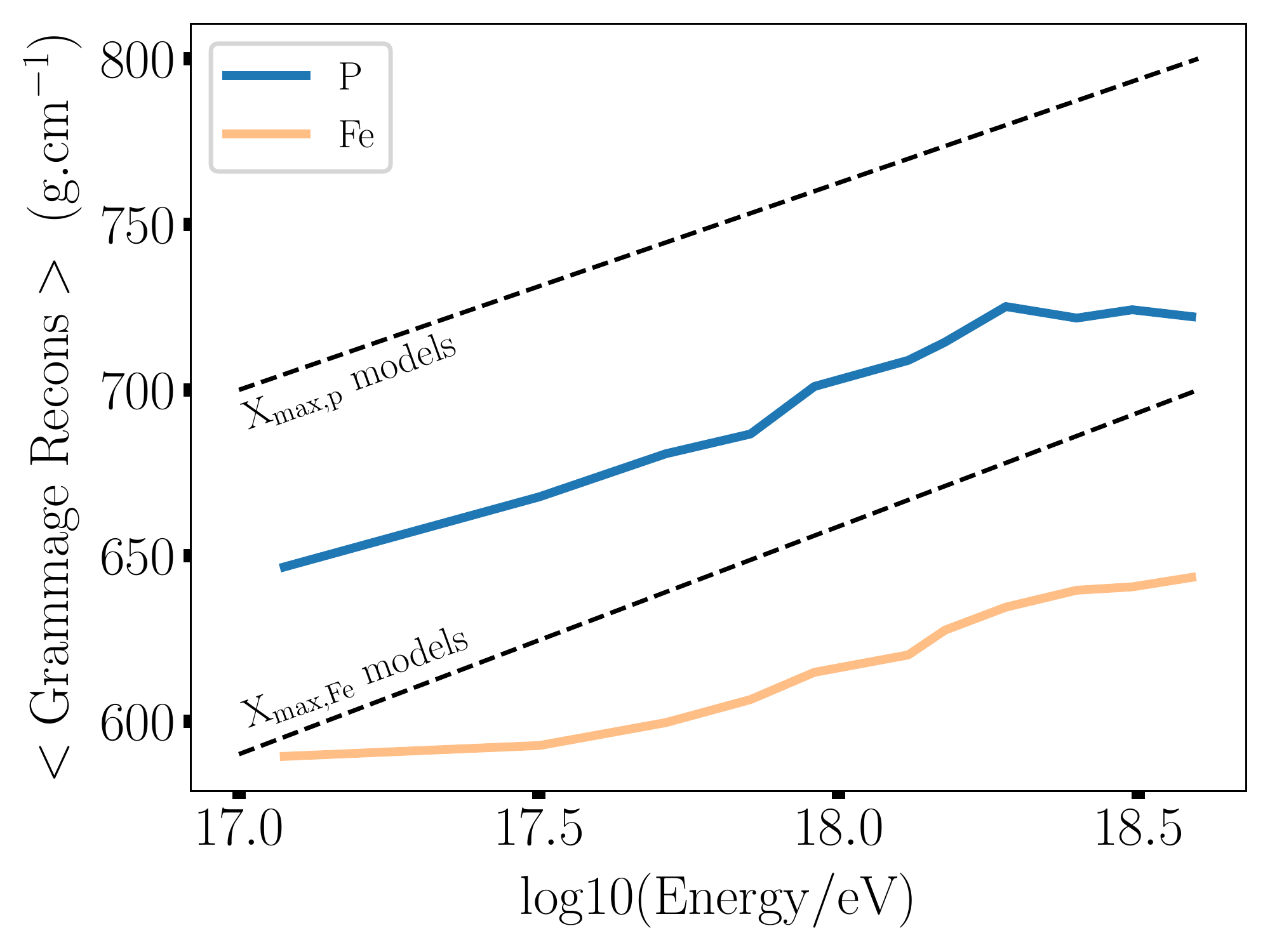} %left
      \includegraphics[width=0.32\linewidth]{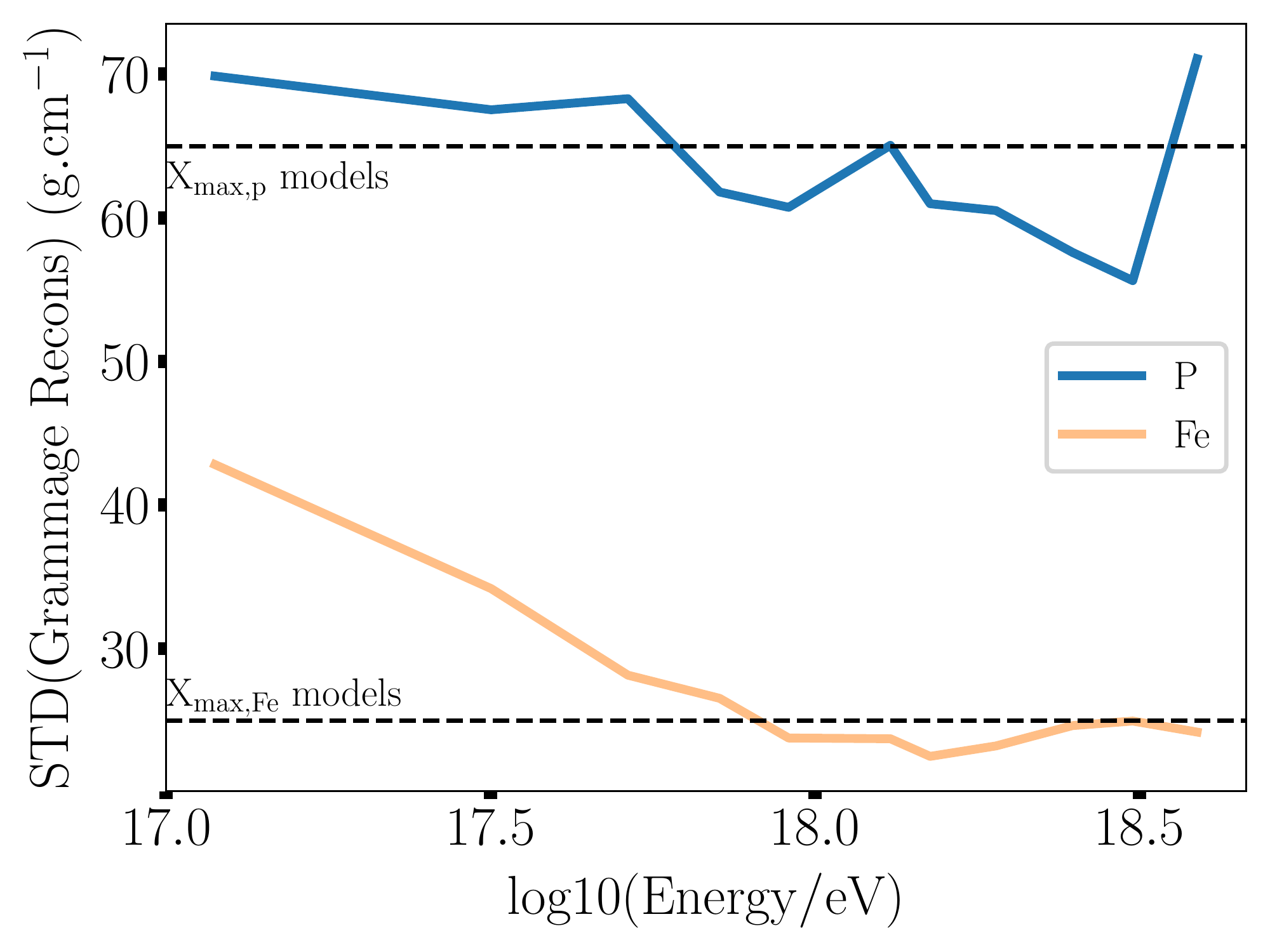} %right
       \includegraphics[width=0.32\linewidth]{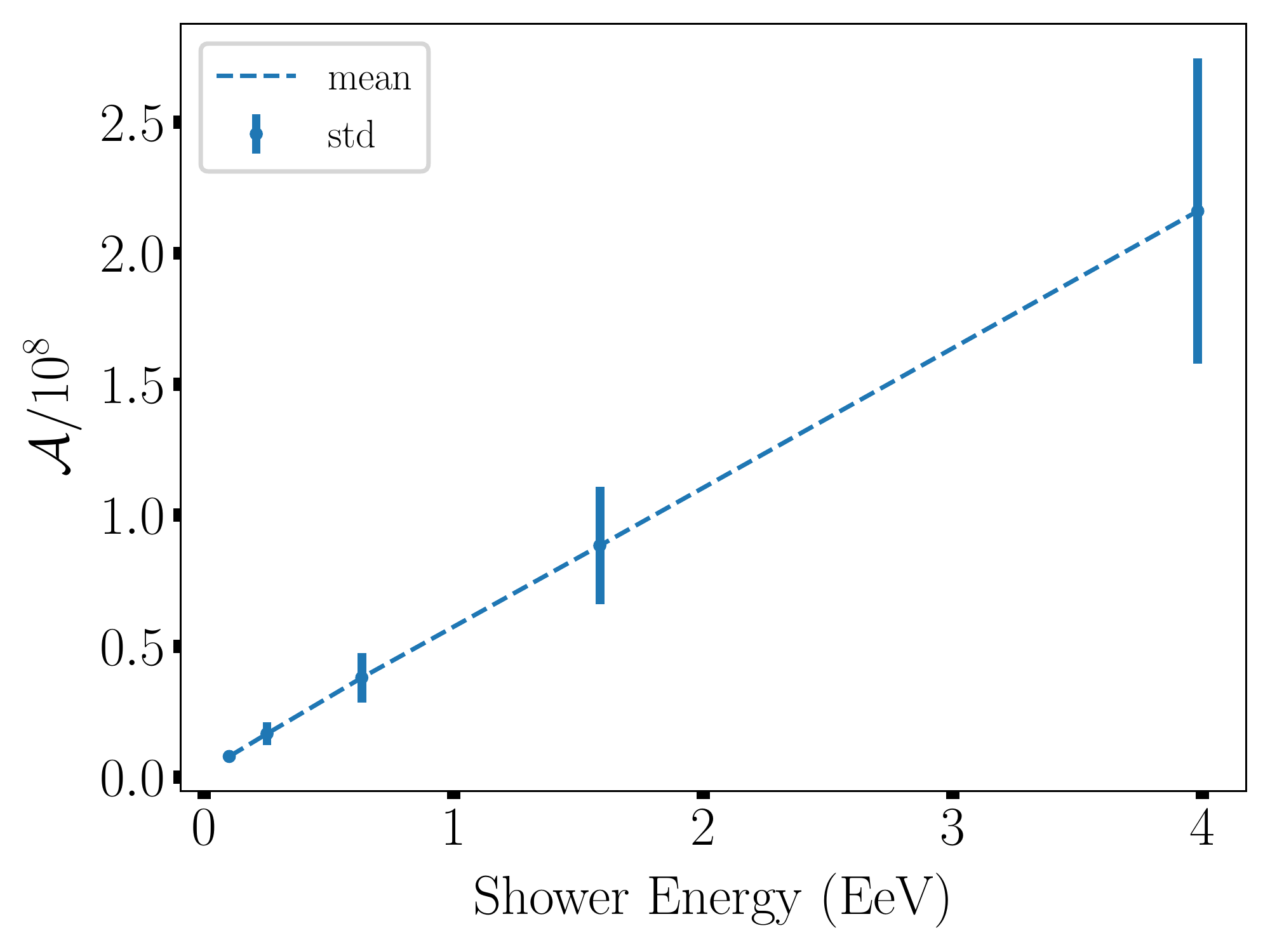}   
     \caption{{\it Left:} The average reconstructed grammage as a function of energy for Proton and Iron primaries. {\it Middle:} The standard deviation of the reconstructed grammage as a function of energy for Proton and Iron primaries. {\it Right:} Energy dependancy on the scaling factor $\mathcal{A}$ of the ADF reconstruction. Results are obtained on star-shape benchmark simulations, see the text.}
   \label{fig:GP300_grammage_energy}
\end{figure}

From the emission point reconstruction it is straightforward to derive the equivalent grammage of that point by integrating over the atmosphere column density along the reconstructed trajectory down to the reconstructed emission point. This effective grammage provides an interesting proxy in the determination of the primary particle composition, as shown in Fig.~\ref{fig:GP300_grammage_energy} for Proton and Iron primaries. Typical resolution of around $60-70$\,g.cm$^{-3}$ and $25-40$\,g.cm$^{-3}$ are obtained for Proton and Iron simulations respectively.
Finally the right panel of Fig.~\ref{fig:GP300_grammage_energy} displays the correlation that can be drawn from the $\mathcal{A}$ amplitude parameter fitted from the ADF model and the total shower energy. The linear correlation and the quite reasonable dispersion observed are promising but remain to be studied in more details. In particular, a lower dispersion is expected by looking at the electromagnetic energy instead of the primary energy.

%%

%----------------------------------------------------------------------
\section{Conclusion and perspectives}
\label{sec:conclusion}
A complete reconstruction procedure for near horizon EAS has been presented. It is based on the combination of information from the arrival times and the amplitudes of the radio signals, in a step by step reconstruction method. The arrival times allow for the construction of an emission point thanks to a spherical wavefront description, which can then be used for the determination of the shower direction thanks to the fitting of the amplitude pattern with the ADF model. The performances achieved on simulations for a realistic radio array in experimental conditions (here GP300) provide an accuracy on the arrival direction within $0.1\degree$, significantly better than what can be achieved currently in similar conditions. Furthermore the determination of the emission point and direction allows for the computation of the emission point equivalent grammage, a new proxy proved to be potentially competitive in the identification of the mass composition of the primaries. Finally, the correlation of the free parameter $\mathcal{A}$ from the ADF fit with the electromagnetic energy of the shower shows promising possibilities in the reconstruction of the energy of the primaries.

Future developments are required to improve the ADF method regarding the fit and optimisation of the reconstruction.

In a broader perspective this study explores the combination of different scalar quantities, measured from the EAS radio signal, in a same reconstruction procedure. Such approach could be generalised to a global reconstruction procedure where all scalar quantities are combined at once, or even use a model that describes the electric field vector in a complete reconstruction procedure, hence minimising the information losses when going from a vector quantity to a scalar one.

\end{document}